# Kinetics of transformation, border of metastable miscibility gap in Fe-Cr alloy and limit of Cr solubility in iron at 858 K

S. M. Dubiel[1,2*] and J. Żukrowski[2]

[1]AGH University of Science and Technology, Faculty of Physics and Applied Computer Science, al. A. Mickiewicza 30, 30-059 Kraków, Poland, [2]AGH University of Science and Technology, Academic Center for Materials and Nanotechnology, al. A. Mickiewicza 30, 30-059 Kraków, Poland

## Abstract

The study was aimed at determination of the position of the Fe-rich border of the metastable miscibility gap (MMG) and of the solubility limit of Cr in iron at 858 K. Towards this end a $Fe_{73.7}Cr_{26.3}$ alloy was isothermally annealed at 858 K in vacuum up to 8144 hours and Mössbauer spectra were recorded at room temperature after every step of the annealing. Three spectral parameters viz. the average hyperfine field, $<B>$, the average isomer shift, $<IS>$, and the probability of the atomic configuration with no Cr atoms in the two-shell vicinity of the probe Fe atoms, $P(0,0)$, gave evidence that the transformation process takes place in two stages. All three parameters could have been well described in terms of the Johnson-Mehl-Avrami-Kolmogorov equation, yielding kinetics parameters. The first stage, associated with the phase decomposition, proceeded much faster than the second stage, associated with the alpha-to-sigma phase transformation. The most reliable estimation of the position of the MMG and that of the value of the Cr solubility limit was obtained from the annealing time dependence of $<B>$, namely 24.5 at.% Cr for the former and 20.3 at.% Cr for the latter. A comparison of these figures with the recent phase diagrams pertinent to Fe-Cr system was done.

Key words: Fe-Cr alloys; phase diagram; miscibility gap; solubility limit; Mössbauer spectroscopy

# 1. Introduction

Fe-Cr alloys have been likely the most frequently studied binary alloys. The unusual interest in them follows, on one hand, form their interesting magnetism (ferromagnetism, antiferromagnetism, spin-glass), and on the other hand from their technological importance especially in the steel making industry. Regarding the latter the Fe-Cr alloys make up the major component of stainless steels, among which ferritic/martensitic (FM) ones have been used as important structural materials in numerous sectors of industry e. g. power pants (including nuclear ones), chemical and petrochemical industries. This role of FM steels follows from their excellent properties and, in particular, high toughness, good resistance to high-temperature corrosion and low swelling. Accordingly, they have been used in various industrial branches to produce devices that work at service at elevated temperatures, often in aggressive environment and under neutron irradiation. Concerning the nuclear power plants, for example, their life time is limited by a degradation of structural devices like vessel and primary circuit due to exposure to radiation and high temperature. The former generates radiation damage and the latter thermal aging. Both result in degradation of mechanical properties and enhanced corrosion. The main reasons for the high-temperature degradation are precipitation of: (1) Cr-rich $\alpha'$ phase and (2) $\sigma$-phase. Both effects can explained based on the crystallographic phase diagram of the Fe-Cr system. Following the Fe-Cr phase diagram [1], the maximum temperature at which precipitation of $\alpha'$ occurs is ~900 K. Yet, current *in-situ* neutron diffraction studies performed on a quasi equiatomic Fe-Cr alloy found that the top of the miscibility gap happens rather at 853 K [2]. The $\alpha'$ precipitation causes significant embrittlement that is known in the literature as "475$^o$C embrittlement" due to the fact that the highest embrittlement rate is at 475$^o$C. The content of Cr in $\alpha'$ is greater than ~85 at% what makes it so brittle. The $\sigma$-phase, in turn, is prone to precipitate if the annealing temperature lies between ~770 and ~1100 K, and the content of Cr is between ~15 and ~85 at.%. The $\alpha'$ precipitates as a consequence of the so-called phase decomposition that results in formation of Fe-rich ($\alpha$) and Cr-rich ($\alpha'$) phases. An interest in this phenomenon has two aspects. First, one wants to know mechanism(s) underlying the decomposition process, and, second, borders of the field in which it occurs (so-called miscibility gap - MG). Two mechanisms have been suggested based on many studies: (1) nucleation and growth (NG), and (2) spinodal (SP). The latter is active in the central part of MG, while the former on both "sides" of the SP field [2]. Thus, the phase fields of NG are close to the Fe-rich and Cr-rich borders of MG. Fe-rich border of MG is more important from the technological viewpoint because it is located close to the Cr concentration at which the Fe-Cr alloy becomes stainless i.e. ~10.5 at%. This Fe-rich border line can be also regarded as the limit of Cr solubility in iron. Despite numerous studies, both theoretical and experimental, were dedicated to this matter clear cut picture has not emerged yet. Theoretical calculations give different predictions, e. g. [1, 3-10], and experimental data, obtained with different techniques like: transmission electron microscopy, electron dispersive microscopy, Mössbauer spectroscopy, small angle neutron scattering, diffuse neutron scattering and resistivity measurements, have a broad dispersion [6]. The dispersion can be, on one hand, due to diverse sensitivity and special resolution of various methods applied, and, on the other hand, it can be caused by different conditions under which samples were investigated (size of samples, their initial degree of homogeneity and strain as well as the annealing time). On the contrary, values of the solubility limit determined by means of the Mössbauer spectroscopy (MS) display systematic trend [11-15]. Thus one can hope that a set of the Mössbauer data constitutes a good basis for validation of different predictions pertinent to the matter in question. In this paper it is reported on the data obtained by studying a $Fe_{73.7}Cr_{26.3}$ sample isothermally annealed at 858 K for up to 8144 hours.

## 2. Experimental

Measurements were performed on a ~25 μm thick foil in form a 20x20 mm rectangle obtained by rolling down ~100 μm thick tape of a $Fe_{73.7}Cr_{26.3}$ alloy. The alloy, in turn, was prepared by melting in an induction furnace under protective Ar atmosphere adequate masses of Armco-iron and chromium of 99.9% purity. The obtained ingot was next rolled down to the thickness of ~100 μm. Its composition was determined by a chemical analysis. To stimulate the decomposition/transformation process, the foil was isothermally annealed under dynamic vacuum (<$10^{-4}$ Torr) at 858 K for up to 8144 hours. After each annealing step, a $^{57}$Fe Mössbauer spectrum was recorded at room temperature in a transmission mode using a standard spectrometer with a drive working in a sinusoidal mode. A $^{57}$Co/Rh source was used as a supplier of 14.4 keV gamma rays. Its activity permitted recording a statistically good spectrum in 1024 channels of a multichannel analyzer within a 2 days run.

The spectra were analyzed assuming that an effect of Cr atoms situated in the first-two neighbor shells around $^{57}$Fe probe nuclei, 1NN-2NN, on the hyperfine field, *B*, and on the center shift, *CS*, was additive i.e. $X(m,n) = X(0,0) + m\Delta X_1 + n\Delta X_2$, where *X=B or CS*, $\Delta X_m$ is a change of *X* caused by one Cr atom present in 1NN (*m*=1) or in 2NN (*m*=2). This procedure has already proved to properly work in the analysis of Mössbauer spectra registered on various Fe-based binary alloys including Fe-Cr ones e. g. [11-20]. The total number of possible atomic configurations (m,n) in the 1NN-2NN approximation is equal to 63. However, for *x* = 26.3 at% most of them have negligible probabilities hence 17 most probable (according to the binomial distribution) were taken into account in the fitting procedure (their overall probability was > 0.97). However, their probabilities, *P(m,n)*, were treated in the fitting routine.as free parameters. As free parameters were also regarded *X(0,0)*, line width (common to all sextets), *G*, $\Delta B_1$, $\Delta B_2$, and an angle between the magnetization vector and normal to the sample's surface, theta. On the other hand, based on our previous studies values of $\Delta CS_1$ = -0.02 mm/s, and $\Delta CS_1$= -0.01 mm/s, were kept constant [18]. Using this procedure all measured spectra could have been well fitted with the average values of: *B(0,0)*=333(2) kGs, $\Delta B_1$=-32.8(5) kGs, $\Delta B_2$=-21.8(3) kGs, *IS(0,0)*=-0.103(3) mm/s, *G*=0.24(1) mm/s.

Examples of the spectra are presented in Fig. 1. The effect of annealing time can be best seen in the outermost lines. Noteworthy is also low intensity of the second and fifth lines in the thermally untreated sample. It is due to a texture induced by cold rolling. Already annealing during 0.25 h has a great effect on the texture. In addition, the spectra were analyzed using a magnetic hyperfine field distribution method to independently and visually illustrate the effect of annealing on the hyperfine field. Selected examples are displayed in Fig. 2.

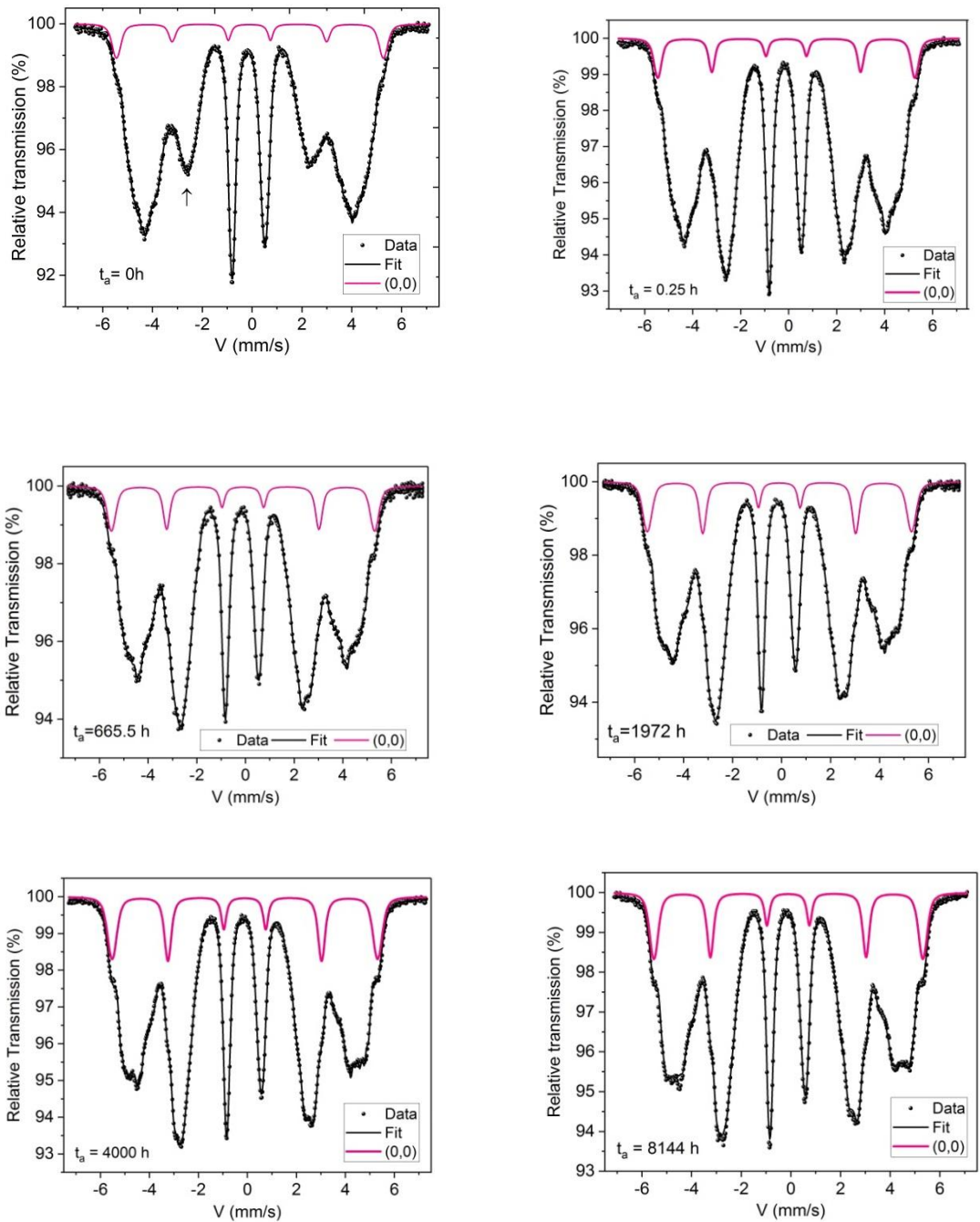

Fig. 1 Examples of Mössbauer spectra recorded on the $Fe_{73.7}Cr_{26.3}$ sample annealed at 585°C for different periods, $t_a$, as indicated for each spectrum.

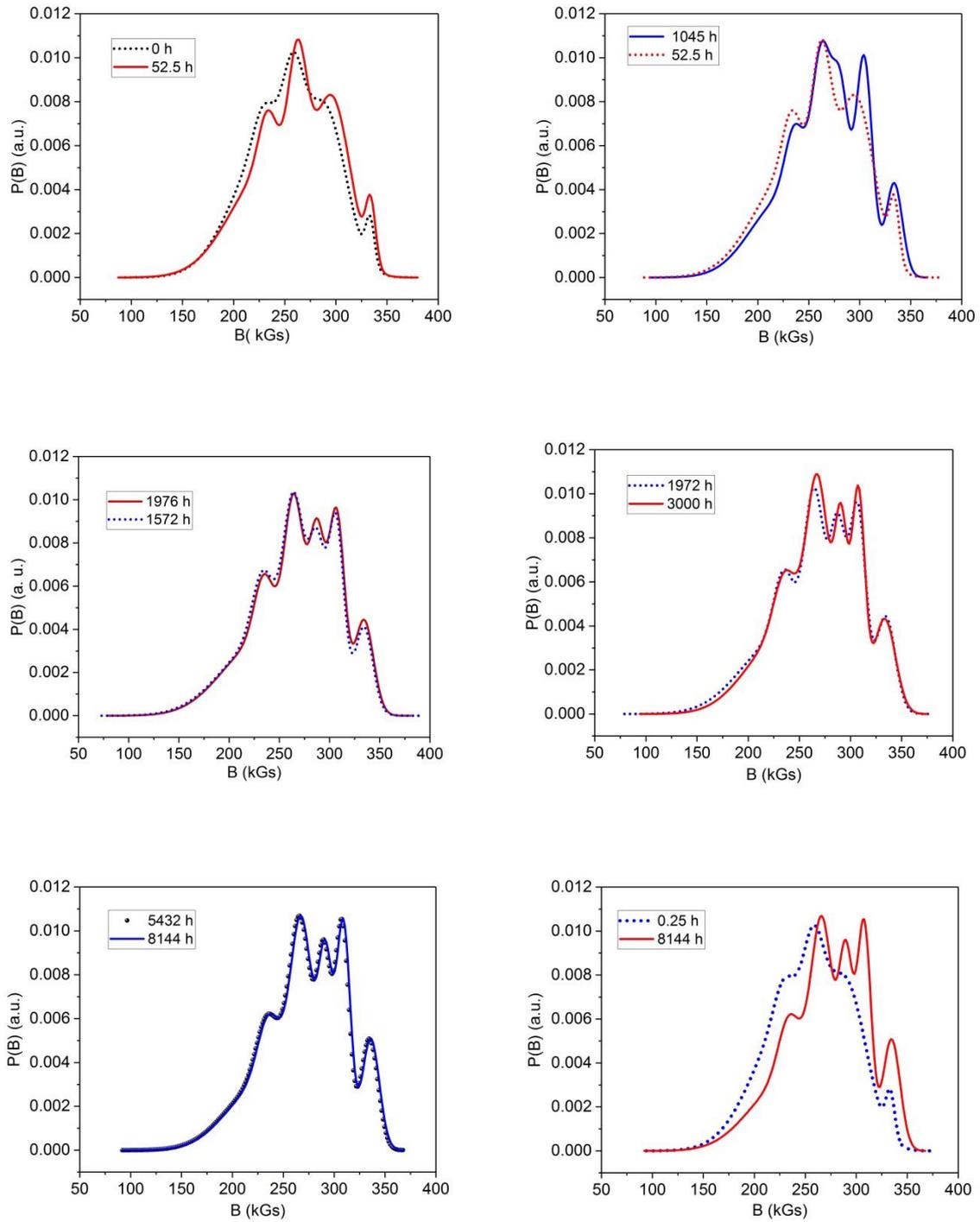

Fig. 2 Comparison between the P(B)-curves derived from the spectra annealed for different periods between 0 h and 8144 h. Notice that there is no more difference between the P(B)-curves for 5432 h and 8144 h.

## 3. Results and discussion

### 3.1. Effect on hyperfine magnetic field

The magnetic hyperfine field, *B*, is the most sensitive spectral parameter to the presence of Cr atoms in the vicinity of the probe Fe atoms. Consequently, a redistribution of Cr atoms caused by the annealing has the most profound effect on *B*. As demonstrated in our previous studies e. g. [15], its annealing time dependence, *B(t)*, can be used to follow a kinetics of transformation, and the maximum value of *B* in a given transformation process, can be further used for determining the concentration of Cr e. g. in the Fe-rich ($\alpha$) and the Cr-rich phases e. g. [15]. The dependence of the average value of *B*, $<B> = \sum B(m,n) \cdot P(m,n)$, on the annealing time, *t*, as found in the present study, is displayed in Fig. 3.

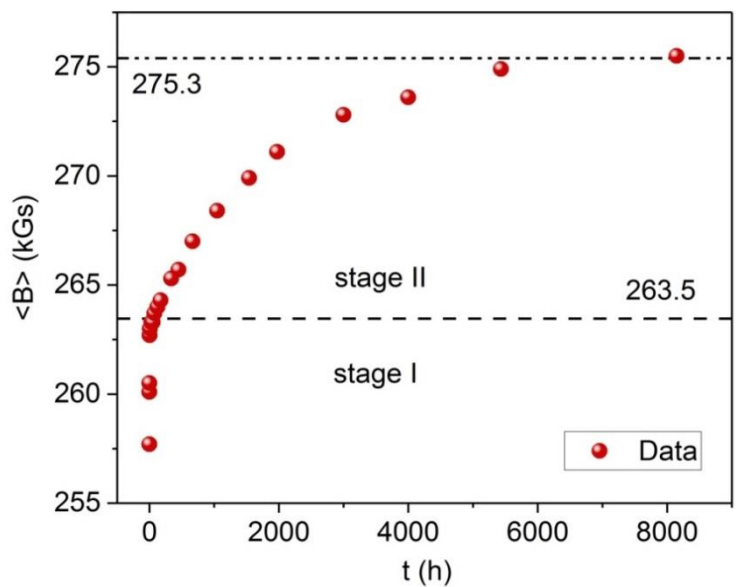

Fig. 3 Average hyperfine field, *<B>*, vs. annealing time, *t*. Two stages of the decomposition process are marked – see also Fig. 4. Maximum values of *<B>* in each of the stages are denoted by horizontal lines and correspondingly labelled.

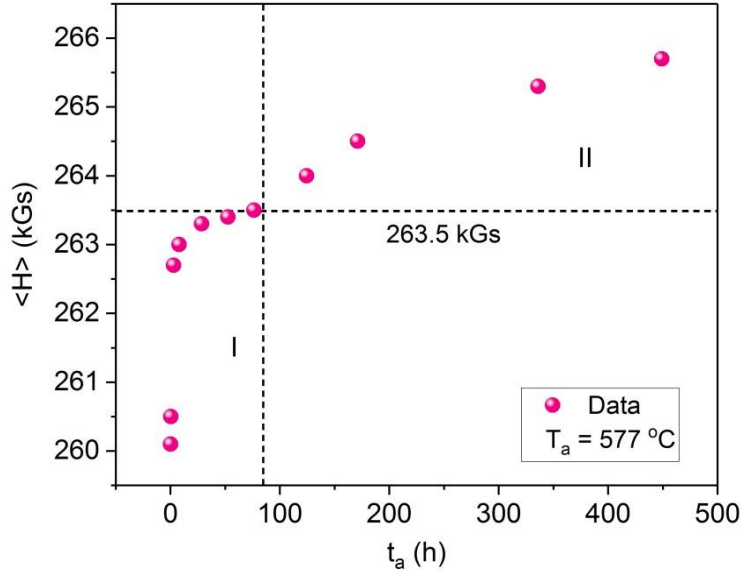

Fig. 4 Average hyperfine field, <B>, vs. annealing time, *t*, for the first ~450 hours of annealing. A vertical dashed line indicates a transition between the first and the second stage of a transformation process. The horizontal dashed line marks the maximum value of <B> in the first stage.

It has been demonstrated in our previous studies that the kinetics of the annealing induced processes could be well described in terms of the Johnson-Mehl-Avrami-Kolmogorov (JMAK) equation [12-15]. In particular, the annealing time dependence of <B> can be expressed like follows:

$$<B>(t) = <B_o> + a \cdot [1 - \exp(1 - (k \cdot t)^n)] \qquad (1)$$

Where <$B_o$> is the value of the average hyperfine field for the non-annealed sample, *k* is the rate constant, *n* is the Avrami exponent, and *a* is a free parameter.

Figure 4 illustrates the <B>(t)-data and its best-fit in terms of eq. (1) for the first stage of the transformation process that can be associated with the phase decomposition. It has terminated after ~80 hours of annealing and the maximum value of <B>=263.5 kGs. This value can be further used for determining a position of the extended miscibility gap line at 858 K. Towards this end can be used a linear relationship between <B> and Cr content, *x*, for binary Fe-Cr [18,19]. By doing so, one arrives at $x$=24.5 at.%. The obtained value of the Avrami exponent, $n$, is equal to 0.5, indicating thereby that the process responsible for the phase decomposition in this sample could be a diffusion-controlled thickening of plates after their edges have impinged i.e. the one dimensional growth of nuclei controlled by diffusion [21].

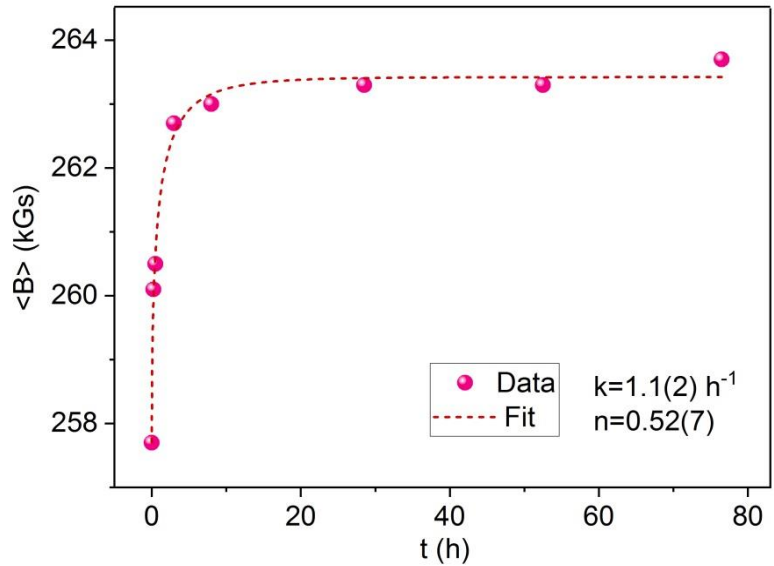

Figure 5. Average hyperfine field, <$B$>, as a function of the annealing time, $t$, for the first stage of the transformation process. The best-fit of the data to eq.(1) is indicated by the dashed line. Values of the kinetic parameters $k$ and $n$ are displayed.

The second stage of the transformation process can be associated with a formation of the σ-phase. Despite long time of annealing, no trace of σ can be seen in the spectrum recorded for the longest period i.e. 8144 h. However, the process of the α to σ transformation is very slugish, hence a very long annealing time is needed [11]. The <$B$>(t) data displayed in Fig. 3 have a saturation-like behavior, so one can assume that the transformation process has alsmost finished after at over 8000 hours of annealing. In other words, the maximum value of <$B$> can be used for estimation of the solubility limit of Cr in iron at 858 K. To find this figure, the <$B$>(t) - data in the second stage of the transformation process i.e. for $t >$ ~80 h were analyzed in terms of eq.(1). The results of this analysis are presented in Fig. 6.

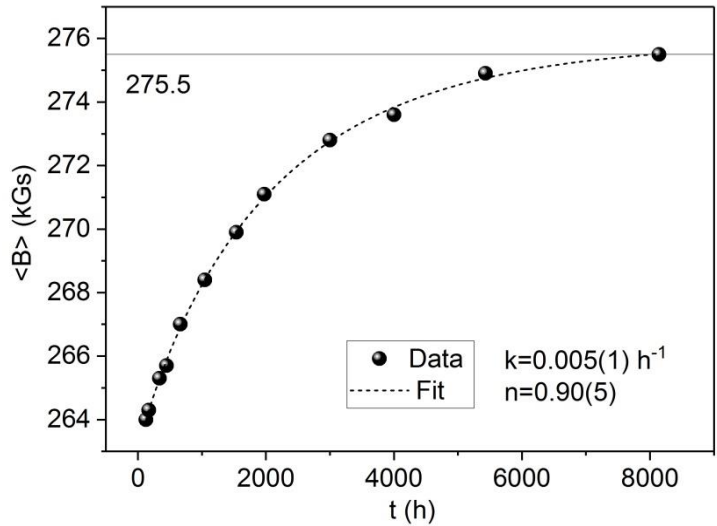

Fig. 6. Average hyperfine field, <B>, as a function of the annealing time, t, for the second stage of the transformation process. The best-fit of the data to eq.(1) is indicated by the dashed line. Values of the kinetic parameters k and n as well as that of <B>$_{max}$ are shown.

Using the same procedure as in the case of the first stage, one finds that the value of <B>=275.5 kGs corresponds to x = 20.3 at.% Cr. It is of interest to compare the kinetics parameters in both stages. Concerning the rate constant, k, the one for the second stage is 200 times smaller than k found for the first stage. This further supports our interpretation that in the first stage, which is very fast, one has to do with the phase decomposition, whereas in the second stage, being two orders of magnitude slower, the process of the phase transformation ($\alpha$ to $\sigma$) dominates. The difference between the two processes has been also reflected in the values of the Avrami exponent, n, which for the second stage of the transformation ~1. Such value of n is characteristic of the diffusion-controlled growth of isolated plates and/or needles of finite size [21]. Such forms are typical of $\sigma$-phase precipitates [22,23].

### 3.1.1. Phase diagram

The value of the Cr solubility limit in iron determined in this study viz. 20.3 at.% can be next used to upgrade the data on the crystallographic phase diagram of the Fe-Cr system. Below a currant full phase diagram and two its currant fragments amended by the present result are shown.

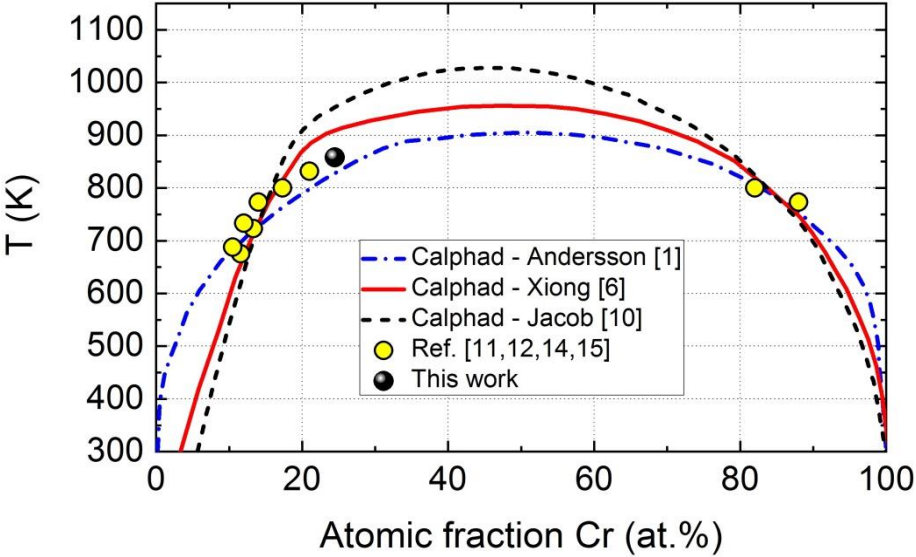

Fig. 7 Comparison between three theoretical Calphad-based predictions concerning the full MG adopted from [10] and our experimental data obtained with the Mössbauer spectroscopy. The size of the corresponding symbols is comparable with the experimental error in the Cr content.

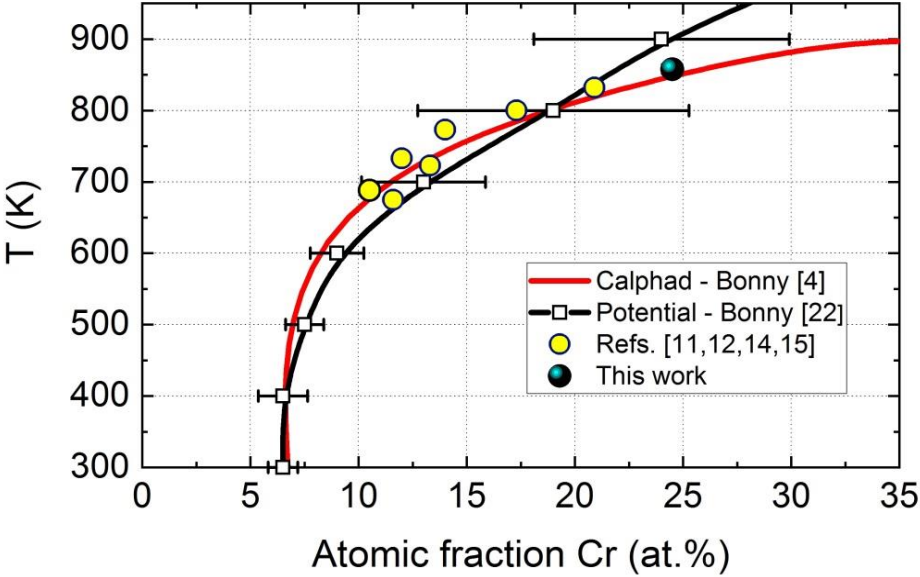

Fig. 8 Comparison between two theoretical predictions depicting the miscibility gap in the Fe-Cr system [24] and experimental data obtained with the Mössbauer spectroscopy. The size of the symbols is comparable with the error in the Cr content..

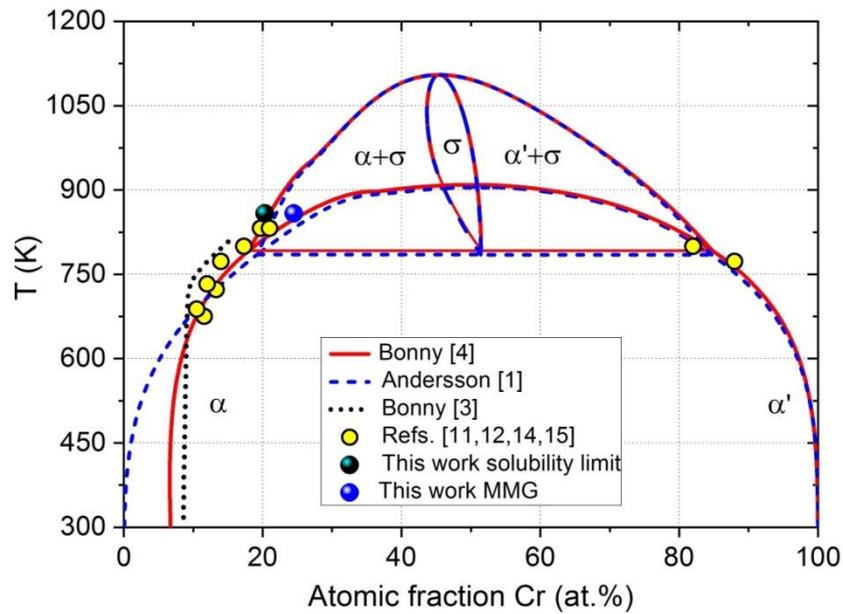

Fig.9 Relation between experimental data obtained with the Mössbauer spectroscopy and the calculated phase diagram of the Fe-Cr system adopted from [4]. Size of the symbols corresponds to the uncertainty in determination of the Cr content.

### 3.2. Effect on isomer shift

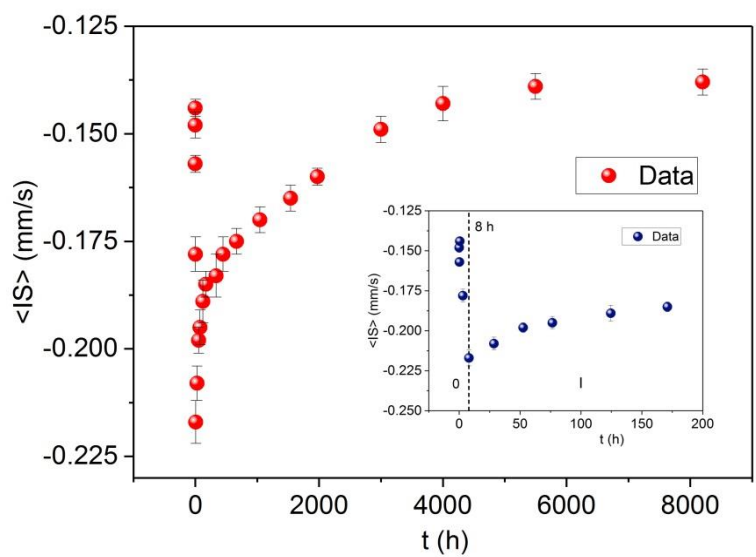

Fig. 10 Annealing time dependence of the average isomer shift, <*IS*>. The inset shows the <*IS*>(t) behavior within the first 170 hours of the annealing.

As illustrated in the inset of Fig. 10, the values of <*IS*> drop sharply within the first 8 hours of annealing. Namely from -0.144 mm/s to -0.217 mm/s i.e. by 0.073 mm/s. Following the relationship between the isomer shift and the density of 4s-like electrons [25] this change of <*IS*> is equivalent to an increase of the charge-density at Fe-nuclei by ~0.07 4s-like electrons. As the investigated sample was initially in a strongly deformed (by cold rolling) state, the first stage of the annealing, denoted in Fig. 2 by "0" can be associated with the removal of strain and related therewith redistribution of Cr atoms [26]. The real process of the transformation starts at $t$≈8h, what is reflected by a continuous increase of the <*IS*>-values. The behavior is quite similar to that of <*B*>(t) – see Fig. 3. The increase of <*IS*> signifies a decrease of the s-like electrons charge-density at the probe atoms. This effect can be understood as due to a decrease of a number of Cr atoms in the vicinity of Fe atoms, a consequence of the phase decomposition into a Fe-rich ($\alpha$) and Cr-rich ($\alpha$') phases [11]. Like in the case of the <*B*)(t), the <*IS*>(t)-data presented in Fig. 10 can be treated as a two-stage process, yet a distinction between them is no as clear-cut as it is in the case of the average hyperfine field. Rather arbitrarily we have made the border between the two stages at 335 h and fitted the two sets of the data to the JAMK-equation. The best-fit curve and the best-fit kinetic parameters for the first stage (phase decomposition) are presented in Fig. 11, whereas those for the second stage (formation of $\sigma$) can be seen in Fig. 12.

Noteworthy, the maximum difference in <*IS*> in both stages is equal to 0.034 mm/s what is equivalent to a decrease of the density of s-like electrons in the conduction band by 3.5% (assuming 1 4s-like electron per atom).

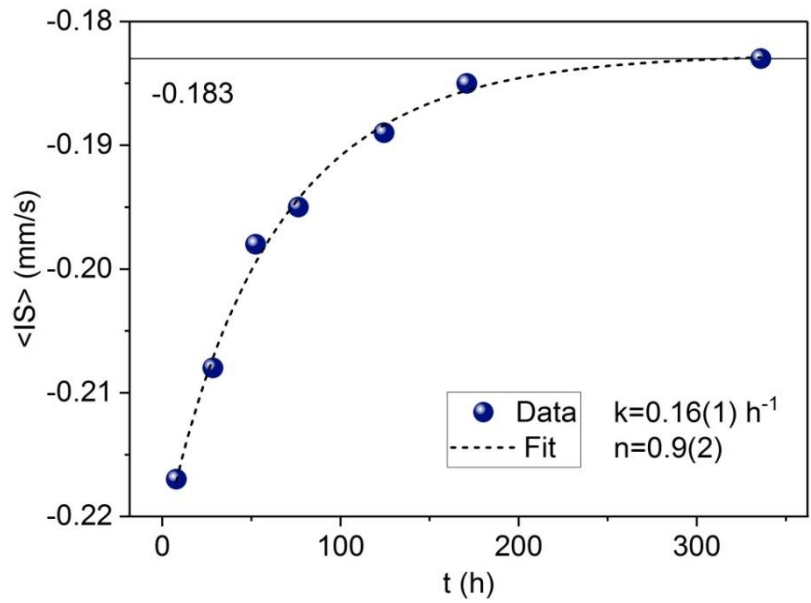

Fig. 11 Annealing time dependence of the average isomer shift, <*IS*>, for the first 335 hours of annealing. The dashed line stands for the best-fit of the data to the JMAK-equation. The best-fit parameters are displayed. The value of <IS> -0.183 mm/s in saturation is indicated.

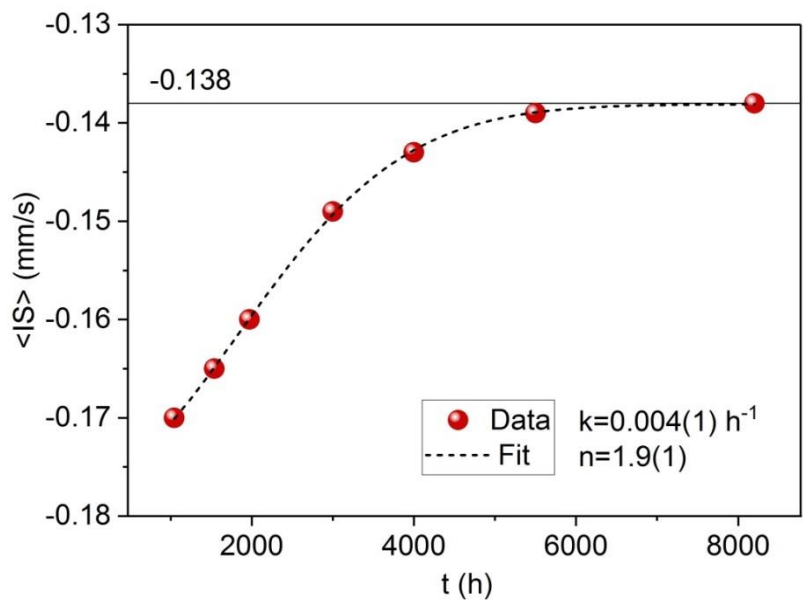

Fig. 12 Time dependence of the average isomer shift, <*IS*>, for the final stage of annealing. The solid line stands for the best-fit of the data to the JMAK-equation. The value of <IS>=-0.138 mm/s in saturation is marked by a horizontal line. The-best fit kinetic parameters are shown, too.

The transformation process shown in Fig. 4 can be associated with the formation of the σ-phase (which is a very slow process). The maximum value of <IS> =-0.138 mm/s can be also used for estimation of the Cr content in the Fe-rich phase. For this purpose we use the linear relationship between <IS> and Cr concentration in binary Fe-Cr alloys [19] and arrive at $x$=19.2 at.% Cr. This figure is in line with the corresponding one determined based on <B>($t$) i.e. 20.3 at.% Cr.

As displayed in Figs. 11 and 12, values of both kinetic parameters are significantly different for the two stages of the transformation processes. In particular, the kinetic parameters, $k$, for the second stage is 40 times smaller indicating thereby that the second stage of transformation is much slower. This is in accord with the fact that the process of α to σ phase transformation is sluggish.

### 3.3 Effect on P(00)

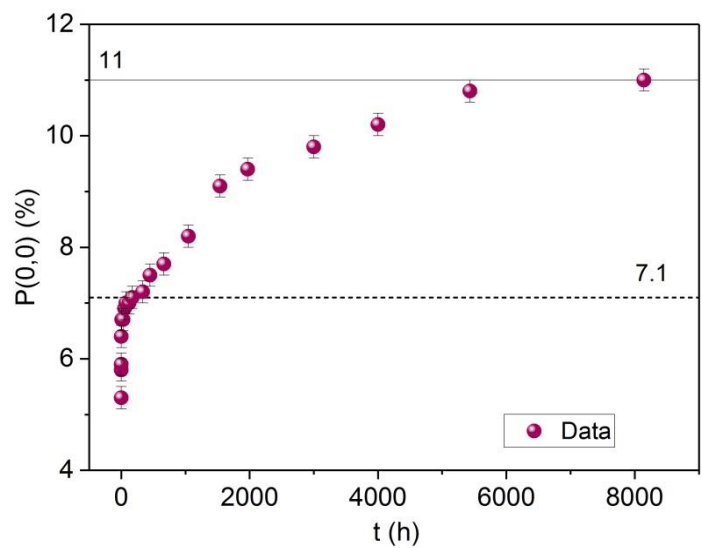

Fig. 13 Dependence of the *P(0,0)* probability on the annealing time, *t*, for the whole range of annealing viz. $0 \leq t \leq 8144$ h. The line at *P(0,0)*=7.1 % stands for the maximum value of *P(0,0)* achieved in the first stage of transformation. The line at $P(0,0)$=11% indicates the maximum value of $P(0,0)$ after the maximum time of the annealing.

The first stage of the transformation process is also visible in the $P(0,0)$(t) behavior. Here it seems to have terminated between ~150 and ~350 hour of annealing – see Fig. 14. The annealing time dependence of $P(0,0)$ can be well described in terms of the JMAK-like equation. The $P(0,0)$-value in saturation, equal to 7.1%, can be used to estimate the concentration of Cr in the Fe-rich matrix, $x$, after the annealing time of < ~350 h. Assuming the distribution is random, $x = 1 - \sqrt[14]{P(0,0)}$ =17.2 at.% i.e. it decreased by 5.9% relative to the Cr content revealed by a chemical analysis. This value is significantly smaller than the corresponding value found from the average hyperfine field viz. 24.5 at.%. Two reasons for the discrepancy can be given: first, the real distribution may deviate from random, and second, the quantity $P(0,0)$ gives information on the local Cr content i.e. from the view of one of many atomic configurations, whereas the information derived from <B> concerns the average concentration.

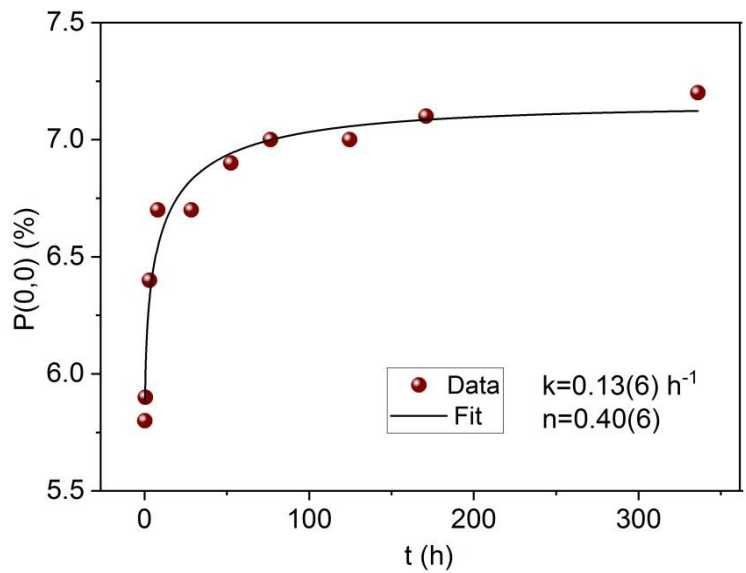

Fig. 14 Dependence of the $P(0,0)$ probability on the annealing time, t, for the first stage viz. $0 \leq t \leq$ ~350 h. The line represents the best-fit of the data to the JMAK-like equation. The kinetics parameters, $k$ and $n$ are displayed.

The P(0,0) data relevant to the annealing period between ~350 h and 8144 h and its best fit in terms of the JMAK-like equation are presented in Fig. 15 alongside with the kinetics parameters $k$ and $n$.

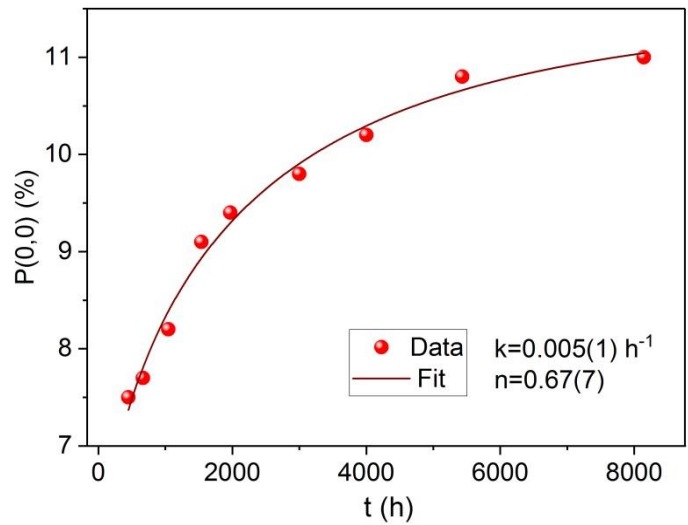

Fig. 15  Dependence of the *P(0,0)* probability on the annealing time, *t*, for the annealing time between ~350h and 8144h. The line represents the best-fit of the data to the JMAK-like equation. The kinetics parameters, *k* and *n* are displayed.

The transformation process or processes in this stage of the annealing is obviously much slower than within the first ~350 h of annealing which is reflected in the value of the rate constant, *k*, being 26-times smaller in the second stage. This again gives evidence that the transformation process in the second stage may be associated with the formation of the σ-phase.

**3. 4 Effect on magnetic texture**

The performed analysis of the Mössbauer spectra also gave information on the magnetic texture. Namely, determined was an angle between the local magnetization vector and the direction of the gamma rays (normal to the sample's surface), theta. Its dependence on the time of annealing, *t*, can be seen in Fig. 16. It can be noticed that theta increases with *t* i.e. the magnetization vector rotates towards the sample's surface, however, as evidenced by the inset, the rate of the increase is not constant but tends to saturate within the first ~125 hours of annealing. This effect seems to be related to the first stage of the transformation. For longer annealing times theta continues to increase and it reaches its maximum of ~60.7 deg at $t \approx 3000$ h. The increase of theta is obviously connected with the growth of grains in the direction parallel to the sample's surface. Then theta gradually decreases approaching the value of ~59.6 deg. The decrease is likely related to the formation of σ that occurs, first of all, on the grains surface and on the sample's surface [27,28]. Consequently, the size of grains shrinks causing thereby a backward rotation of the magnetization vector.

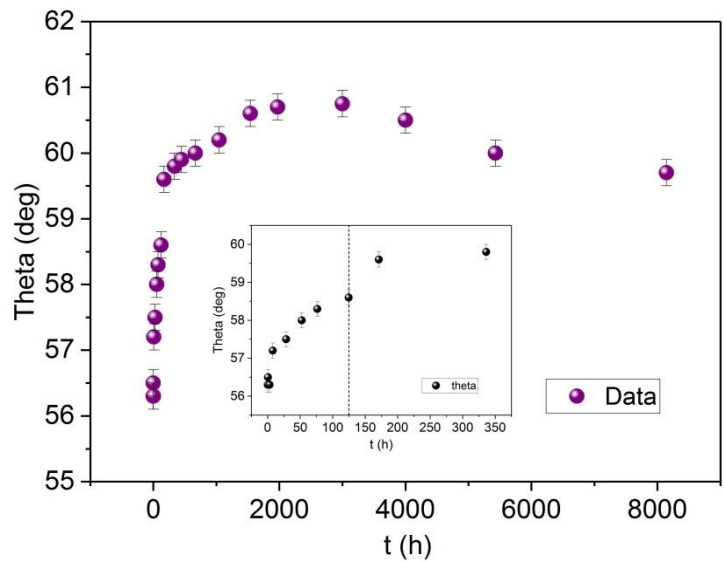

Fig. 16 Dependence of theta on the annealing time, *t*, for the whole range of annealing viz. $0 \leq t \leq 8144$ h. The inset shows the behavior of theta within the first ~350 hours of annealing.

## 4. Summary

The effect of a in-vacuum isothermal annealing at 858 K on a transformation process in a $Fe_{73.7}Cr_{26.3}$ alloy was studied by means of the Mösbauer spectroscopy. Considered were four spectral parameters viz. the average hyperfine field, *<B>*, the average isomer shift, *<IS>*, the probability of the atomic configuration with no Cr atoms within the first two coordination shells around the probe Fe atoms, P(0,0), and the average angle between the local magnetization vector and the direction of the gamma rays (normal to the sample's surface). The first three of them give a clear evidence that the transformation process proceeds in two stages. The first stage has been associated with the phase decomposition and the second one with the transformation of the α- into the σ-phase. The annealing time evolution of all three quantities could be well described in terms of the Johnson-Mehl-Avrami-Kolmogorov equation yielding kinetics parameters i.e. the rate constant, $k$, and the Avrami exponent, $n$. Their values are characteristic of a given stage of transformation, and, as a rule, the value of $k$ for the second stage is up to 200 times smaller than the one for the first stage. This means that the α-to-σ phase transformation is correspondingly slower than the preceding phase decomposition. Furthermore, the in-saturation values of the three quantities enabled determination of: (1) position of the metastable miscibility gap (the first stage), and (2) solubility limit of Cr in iron (the second stage). The most reliable values viz. 24.5 at.% Cr for (1) and 20.3 at.% Cr for (2) were obtained from the <$B$>(t) dependence. These values were compared with recent theoretical predictions pertinent to the Fe-Cr phase diagram.

## Acknowledgements

This work was financed by the Faculty of Physics and Applied Computer Science AGH UST and ACMIN AGH UST statutory tasks within subsidy of Ministry of Science and Higher Education. G. Bonny is gratefully thanked for making available his calculations shown in Figs. 8 and 9.## CRediT author statement

**Stanisław M. Dubiel:** Conceptualization, Sample preparation, Investigation, Validation, Formal analysis, Writing - Original Draft, Visualization: **Jan Żukrowski:** Software, Validation, Investigation, Data Curation.